\documentclass[graybox, envcountchap, authoryear, natbib]{svmult}

\usepackage{mathptmx}       % selects Times Roman as basic font
\usepackage{helvet}         % selects Helvetica as sans-serif font
\usepackage{courier}        % selects Courier as typewriter font
\usepackage{type1cm}        % activate if the above 3 fonts are
\usepackage{makeidx}         % allows index generation
\usepackage{graphicx}        % standard LaTeX graphics tool
\usepackage{multicol}        % used for the two-column index
\usepackage[bottom]{footmisc}% places footnotes at page bottom

\newcommand{\OI}{O\,{\footnotesize I}}

\newcommand{\CII}{C\,{\footnotesize II}}
\newcommand{\CIII}{C\,{\footnotesize III}}
\newcommand{\AlIII}{Al\,{\footnotesize III}}
\newcommand{\AlII}{Al\,{\footnotesize II}}
\newcommand{\CIV}{C\,{\footnotesize IV}}
\newcommand{\SiII}{Si\,{\footnotesize II}}

\newcommand{\SiIV}{Si\,{\footnotesize IV}}
\newcommand{\FeII}{Fe\,{\footnotesize II}}
\newcommand{\MgI}{Mg\,{\footnotesize I}}
\newcommand{\MgII}{Mg\,{\footnotesize II}}
\newcommand{\CaII}{Ca\,{\footnotesize II}}
\newcommand{\NaI}{Na\,{\footnotesize I}}
\newcommand{\HeII}{He\,{\footnotesize II}}
\newcommand{\HeI}{He\,{\footnotesize I}}
\newcommand{\OII}{[O\,{\footnotesize II}]}
\newcommand{\OIII}{[O\,{\footnotesize III}]}
\newcommand{\NII}{[N\,{\footnotesize II}]}

\begin{document}

\title*{Gas Accretion Traced in Absorption in Galaxy Spectroscopy}
\author{Kate H. R. Rubin}
\institute{Kate H. R. Rubin \at San Diego State University, Department
  of Astronomy, San Diego, CA 92182 \email{krubin@sdsu.edu}}
%\and Name of Second Author \at Name, Address of Institute \email{name@email.address}}
\maketitle

%\abstract*{Each chapter should be preceded by an abstract (10--15 lines long) that summarizes the content. The abstract will appear \textit{online} at \url{www.SpringerLink.com} and be available with unrestricted access. This allows unregistered users to read the abstract as a teaser for the complete chapter. As a general rule the abstracts will not appear in the printed version of your book unless it is the style of your particular book or that of the series to which your book belongs.
%Please use the 'starred' version of the new Springer \texttt{abstract} command for typesetting the text of the online abstracts (cf. source file of this chapter template \texttt{abstract}) and include them with the source files of your manuscript. Use the plain \texttt{abstract} command if the abstract is also to appear in the printed version of the book.}

\abstract{The positive velocity shift of absorption transitions
  tracing diffuse material observed in a galaxy spectrum 
  is an unambiguous signature of gas flow toward the host system.
  Spectroscopy probing, e.g., \NaI\ resonance lines in the
  rest-frame optical or \MgII\ and \FeII\ in the
  near-ultraviolet is in principle sensitive to the infall of cool
  material at temperatures $T\sim100-10,000$ K anywhere along the line of sight
  to a galaxy's stellar component.  However, secure detections of
  this redshifted absorption signature have proved challenging to
  obtain due to the ubiquity of cool gas outflows giving rise to
  blueshifted absorption along the same sightlines.  In this chapter,
  we review the \emph{bona fide} detections of this phenomenon.
  Analysis of \NaI\ line profiles has revealed numerous instances of
  redshifted absorption observed toward early-type and/or AGN-host
  galaxies, while spectroscopy of \MgII\ and \FeII\ has provided
  evidence for ongoing gas accretion onto $>5\%$ of luminous,
  star-forming galaxies at $z\sim0.5-1$.  We then discuss the
  potentially ground-breaking benefits of future efforts to improve the spectral
  resolution of such studies, and to leverage spatially-resolved
  spectroscopy for new constraints on inflowing gas morphology.  }

\section{Introduction}
\label{sec:1}
Spectroscopy of galaxy continua has been used for decades as a
powerful probe of the kinematics of gas in the foreground.  Absorption
transitions sensitive to cool, diffuse material trace its motion with
respect to the galaxy's stellar component along the line of sight.
The galaxy continuum ``beam'' is absorbed by gas over a
broad range of scales, from the $\sim100-200$ pc scale heights of the
dense interstellar medium \citep{langer14} to the $> 100$ kpc
extent of the diffuse gas reservoir filling the galaxy halo
\citep{prochaska11,tumlinson11,werk14} and beyond.  

The detection of absorption lines in a galaxy's spectrum (i.e., ``down
the barrel'') which are \emph{redshifted} with
respect to its rest frame is unequivocal evidence of
gas flow toward the continuum source.  And while this signature
arises from material at any location along the sightline to the galaxy,
such that the flow may not ultimately reach the galaxy itself,
redshifted absorption has been broadly interpreted as strong evidence
for gas inflow toward or accretion onto the background host.  Over the
past ten years, as high signal-to-noise (S/N) galaxy
spectroscopy covering the rest-frame optical into the near-ultraviolet
has become more routine,  this
technique has become sensitive to the inflow of material over a broad
range of densities and temperatures: redshifted \CaII\ H \& K $\lambda
\lambda 3934, 3969$ or \NaI\ D $\lambda \lambda 5891, 5897$ absorption
probes cold, mostly neutral 
gas infall at a temperature $T < 1000$ K; redshifts in low-ionization transitions such as
\MgII\ $\lambda \lambda 2796, 2803$ or \FeII\ $\lambda \lambda 2586, 2600$
trace the inflow of cool, photoionized gas at $T\sim10^4$ K; and the
detection of redshifted \CIV\ $\lambda \lambda 1548, 1550$ or \SiIV\ $1394,
1402$ absorption would in principle trace yet warmer ($T\sim10^5$ K) gas accretion.

Furthermore, unlike background QSO absorption line experiments, 
%(discussed in Chapter ??) 
which typically must adopt the assumption that
inflowing gas has a relatively low metallicity 
\citep[$Z/Z_{\odot}<1$;][]{lehner13}
in order to disentangle accreting systems from those
enriched by galactic outflows, redshifted self-absorption naturally
traces metal-rich inflow, and may even trace pristine inflow in rare
cases where spectral coverage of Ly$\alpha$ is available \citep{fath16}.
Moreover, while studies searching for the signature of gas accretion
in emission from the 21cm transition of neutral hydrogen 
%(Chapter ??)
are currently limited by the faint surface brightness of such features to
galaxies within a few hundred Mpc of our own \citep{martin10}, the
``down the barrel'' technique has been used to detect gas accretion onto
galaxies as distant as $z\sim1$ \citep{coil11,rubin12,martin12}.

In spite of these clear advantages, however, the first report of
redshifted absorption observed down the barrel toward a sample of
more than a single object did not occur until 2009 \citep{sato09}.
Even today, secure detections of this phenomenon have been reported for 
only $\sim 60-80$ systems.  Instead, measurements of
\emph{blueshifted} absorption tracing cool gas outflow have dominated
the literature
\citep{heckman00,martin05,rupke05b,weiner09,steidel10,chen10}.  
Large-scale galactic winds, thought to be driven by processes associated with star
formation, are now known to arise ubiquitously in star-forming
objects from the local universe to $z>2$ 
\citep{heckman00,ajiki02,martin12,rubin14}.  These winds are
likewise traced by all of the metal-line absorption features listed above, and are
observed to have velocities ranging from $\sim -50~\rm km~s^{-1}$ to
$< -800~\rm km~s^{-1}$ \citep{steidel10,rubin14,du16}.  As
the free-fall
velocity of material within the virial radius of a typical massive,
star-forming galaxy halo at $z\sim0$ (with halo mass $M_h \sim 10^{11-12}M_{\odot}$) is expected
to be only $\sim 100-200~\rm km~s^{-1}$ \citep{goerdt15},  
the preponderance of winds covering the sightlines to galaxies must necessarily
obscure the detection of material falling inward at such comparatively
modest velocities.  Indeed, this issue is compounded by the low
spectral resolution of the vast majority of spectroscopic surveys
useful for these analyses \citep{weiner09,steidel10,martin12,rubin14}.

In this chapter, we review the works in which the few {\it bona
  fide} instances of gas inflow were reported, beginning with the
 first detections via transitions in the rest-frame optical in
Section~\ref{sec:2}.  Due to technical limitations (described below),
the focus of these early studies was on red, early-type galaxies and/or
galaxies exhibiting signs of AGN activity.  Reports of inflow onto
galaxies hosting the most luminous AGN (i.e., bright QSOs) are
discussed in Section~\ref{subsec:qsos}.
In Section~\ref{sec:uv}, we describe the first detections of inflow onto actively star-forming systems
facilitated by  high-S/N spectroscopic
galaxy surveys in the rest-frame ultraviolet.
%I discuss reports of inflow onto galaxies hosting bright QSOs in Section~\ref{subsec:qsos}.
The biases inherent in the use of
the ``down the barrel'' technique given the ubiquity of galactic winds are described in
Section~\ref{subsec:resolution}, and 
unique constraints on the
morphology of gas inflow which will soon be facilitated by ongoing
spatially-resolved spectroscopic surveys are discussed in
Section~\ref{subsec:ifus}.  
Section~\ref{sec:future} presents a summary, and offers some recommendations for future
experiments which will leverage the full potential of this technique
in the detection and characterization of the process of gas accretion
onto galaxies.

%Use the template \emph{chapter.tex} together with the Springer document class SVMono (monograph-type books) or SVMult (edited books) to style the various elements of your chapter content in the Springer layout.

%Instead of simply listing headings of different levels we recommend to
%let every heading be followed by at least a short passage of text.
%Further on please use the \LaTeX\ automatism for all your
%cross-references and citations. And please note that the first line of
%text that follows a heading is not indented, whereas the first lines of
%all subsequent paragraphs are.

%\section{First Detections: Revealing Ongoing Gas Accretion in the Late Stages of Galaxy Evolution}
\section{First Detections: Gas Accretion in Late-Stage Galaxy Evolution}
\label{sec:2}
% Always give a unique label
% and use \ref{<label>} for cross-references
% and \citep{<label>} for bibliographic references
% use \sectionmark{}
% to alter or adjust the section heading in the running head

For galaxy spectroscopy to successfully constrain the kinematics of
absorption lines along the line of sight, it must have sufficient continuum
S/N in the spectral range surrounding the transition in
question.  The limiting S/N depends in detail on a number
of factors, including the typical
equivalent width of the transition and the complexity of the absorption line
analysis  employed.  Studies testing this limit and adopting
a simple, ``single component'' analysis have
required between $\rm S/N \sim 5~\AA^{-1}$ 
\citep[for analysis of \MgII;][]{rubin14} and $\rm S/N \sim 15~\AA^{-1}$ 
\citep[for analysis of \NaI\ D;][]{sato09}.  

Many of the first works to attempt constraints on the
kinematics of such absorption transitions 
\emph{in galaxies not specifically selected to be starbursts} -- i.e.,
in magnitude-limited galaxy samples representative of ``typical'' star-forming or
quiescent objects -- used datasets which did not broadly satisfy these S/N
requirements \citep{weiner09,rubin10,chen10,steidel10}.  Each of these
studies were leveraging spectroscopic samples obtained for the primary
purpose of conducting a redshift survey, rather than an assessment of
cool gas kinematics.  As a result, these works often resorted to co-adding
subsamples of tens or hundreds of spectra to achieve the S/N required for
absorption line analysis.  Moreover, this co-added spectroscopy
yielded absorption profiles which were universally asymmetric with
excess equivalent width blueward of the galaxies' rest frame.  
%It was only through high-S/N spectroscopy of
%\emph{individual} galaxies that 
The first detections of
redshifted absorption profiles were reported only in later studies
which obtained high-S/N spectroscopy of \emph{individual} galaxies.

\subsection{First Reports of Inflow Observed Down the Barrel}
\label{subsec:optical}

The very first detections of redshifted self-absorption %observed ``down the barrel'' 
made use of some of the highest-S/N spectroscopy obtained during the execution
of the DEEP2 redshift survey \citep{davis03}.
In their discussion of a massive ultraluminous
infrared galaxy
with an X-ray-bright central AGN at $z=1.15$ located in the AEGIS survey
field \citep{davis07}, \cite{lefloch07} commented on strong \CaII\ H \& K absorption lines
which were redshifted by $\sim150-200~\rm km~s^{-1}$ relative to the
velocity of \OII\ emission from the host.

Shortly thereafter, \cite{sato09} significantly expanded the sample of
redshifted self-absorption detections through their analysis of \NaI\ D
kinematics in a S/N-limited subsample of the DEEP2 survey of
 the Extended Groth Strip.  The parent DEEP2 sample was magnitude-selected to $R <
24.1$, such that it contained substantial populations of both
star-forming and passive galaxies.  However, \cite{sato09} found that
they required at least $\rm S/N \sim 6.5~ pix^{-1}$ in the rest-frame continuum
around \NaI\ D  to constrain its velocity to within a 68\%
confidence interval spanning less than $\sim200~\rm km~s^{-1}$.  This
limited their analysis sample to 205 objects at $0.11 < z < 0.54$,
about $75\%$ of which are ``red sequence'' galaxies.  
Outflows and inflows are reported in this work if the
fitted \NaI\
centroid is shifted from systemic velocity by more than $\pm50~\rm
km~s^{-1}$, 
and occur in nearly equal
numbers: outflows are detected in 32 objects, while inflows are
detected with high confidence in 31
objects.

\begin{figure}[]
%\sidecaption
% For example, with the graphicx style use
\includegraphics[scale=2.9]{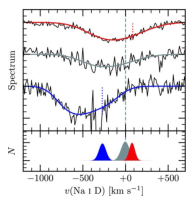}
\includegraphics[scale=2.0]{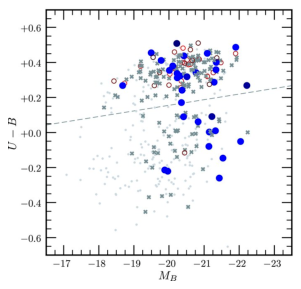}
%\includegraphics[scale=2.9]{Sato09_apj300430f4_hr.jpg}
%\includegraphics[scale=2.0]{Sato09_apj300430f10_hr.jpg}
%
% If no graphics program available, insert a blank space i.e. use
%\picplace{5cm}{2cm} % Give the correct figure height and width in cm
%
\caption{\textbf{Spectroscopy and photometry of the sample analyzed in
  \cite{sato09}, the first study to report and characterize gas
  accretion onto distant galaxies.}  \textit{Left:}
Continuum-normalized spectra showing the region around the \NaI\ doublet
(black).  $v = 0~\rm km~s^{-1}$ is set at the systemic velocity of the
$\lambda 5897$ doublet transition.
The colored curves show the best-fit absorption line models,
and the vertical dotted lines show the corresponding best-fit velocity
of the $\lambda 5897$ line.  The bottom panel shows the marginalized probability
distributions of the model profile velocity for each spectrum,
obtained from a Markov Chain Monte Carlo sampling of the absorption
line profile parameters.
\textit{Right:}  Rest-frame color-magnitude diagram showing the full
sample.  Objects exhibiting outflows are marked with blue filled
circles; those exhibiting redshifted absorption are marked with red
open circles; and gray crosses mark objects which exhibit neither
significant blueshifts nor redshifts in their \NaI\ profiles.  
The light gray dots indicate objects whose spectra lack the S/N
required to constrain their absorption line kinematics. 
The dashed line divides the blue cloud and red
sequence populations.  
Panels are reproductions of Figures 4 (left) and 10 (right) from the
article ``AEGIS: The Nature of the Host Galaxies of Low-Ionization
Outflows at $z<0.6$'', by T. Sato et al.\ (2009, ApJ, 696, 214).  \textcopyright AAS. Reproduced with permission.}
\label{fig:1}       % Give a unique label
\end{figure}

The \NaI\ profiles shown in the left-hand panel of Fig.~\ref{fig:1}
demonstrate the typical quality of the data used in this analysis.  
As the two transitions in the
\NaI\ D doublet are separated by only $\sim300\rm ~km~s^{-1}$, the lines are
blended in these spectra due to the large intrinsic velocity
of both the stellar and gas components giving rise to the observed
absorption.  The velocity $v(\rm \NaI~D) = 0~\rm km~s^{-1}$ on the
x-axis corresponds
to the rest wavelength of the 5897 \AA\ doublet transition.  The
authors show their best-fit absorption line model for each spectrum
with colored curves, and mark the corresponding best-fit velocity of
the $\lambda 5897$ \AA\ transition with vertical dotted lines.  The
top-most spectrum exhibits significantly redshifted absorption with a fitted velocity
of $\sim+80~\rm km~s^{-1}$.  

\cite{sato09} note that while a significant fraction of the galaxies
in their sample which exhibit outflows are star-forming and lie in the
so-called ``blue cloud'' (blue symbols; Fig.~\ref{fig:1}, right), all but one of the
objects exhibiting inflows occupy the red sequence (red symbols;
Fig.~\ref{fig:1}, right).  In interpreting this finding, they caution
that there may be additional absorption features arising from the
stellar populations in these systems which are not yet understood, 
and which could in principle shift the minimum of the stellar
continuum flux near $\lambda_{\rm obs} \sim 5890-5900$ \AA\ redward of
the rest wavelengths of \NaI\ D.  Line-profile fitting of such features could
in turn be mistakenly interpreted as redshifted interstellar absorption.
%possibly yielding a redshift in their
%single-component absorption line fit.  
However, the authors also report that many of the inflow galaxies in their sample
exhibit optical emission line ratios (\NII/H$\alpha$) consistent with
Seyfert or LINER activity, and draw an intriguing comparison to the inflows observed in neutral
hydrogen in the inner regions
of radio-bright elliptical galaxies 
\citep{vangorkom89}.
They speculate that the observed absorbing gas
may in fact feed this central activity. 

\subsection{Inflows onto AGN-Host Galaxies}
\label{subsec:agn}

Since the publication of this novel and important work, there have
been several studies corroborating the detection of inflow toward
galaxies hosting active AGN.  \cite{krug10} explored \NaI\ D kinematics
in a sample of Seyfert galaxies selected to be infrared-faint (with
$10^{9.9} < L_{\rm IR}/L_{\odot} < 10^{11}$).  Previous studies of \NaI\
D absorption toward nearby galaxy samples had targeted infrared-bright
starburst or starburst/AGN composite systems
\citep{rupke05b,rupke05c,martin05}, and had detected outflows in the majority
of these objects.  A primary goal of \cite{krug10} was to develop a
comparison sample of objects without ongoing starbursts 
 to determine the relative contributions of star
formation vs.\ AGN activity in driving these winds.   The authors obtained
spectroscopy of 35 galaxies using the RC Spectrograph on the Kitt Peak
4m telescope.  With a median S/N near \NaI\ D of $\sim85~\rm \AA^{-1}$
and a spectral resolution of $85~\rm km~s^{-1}$, they performed both
single- and double-velocity component absorption line model fits with
typical central velocity uncertainties of $< 50~\rm km~s^{-1}$.  

In contrast to previous studies of starbursting systems, \cite{krug10} 
detected outflows traced by \NaI\ in only  4 galaxies (11\% of their sample)
and instead detected inflows in over a third (13 galaxies) of their sample.  
The central velocities of these flows ranged from just over 
$+50~\rm km~s^{-1}$ to $+140~\rm km~s^{-1}$; however, no significant
correlation between inflow kinematics and host galaxy mass, infrared
luminosity, or inclination was observed.
From this analysis, the
authors concluded that star formation rather than
AGN activity makes the dominant contribution to the driving of outflows
\citep[although this conclusion does not apply to Type 1 Seyferts;][]{krug10}. 
Furthermore, they speculated that the high observed 
inflow velocities are suggestive of material located close to the
galaxy nuclei rather than in the outskirts of the disks.  
They also searched for signs of nuclear morphological features which
could indicate that the inflowing material is undergoing angular momentum loss,
finding that 5 galaxies in their inflow sample exhibit nuclear dust
spirals, bars, or rings.  Five additional inflow galaxies have nearby
companions with which they may be interacting.
Such phenomena are thought to be required in
order to facilitate the inflow of gas toward galaxy nuclei and its
ultimate accretion onto the central black hole.  

The high quality of these data also permitted a rough estimate of the
rate of mass inflow onto these systems via constraints on the \NaI\
column density.  Assuming a factor of 10 ionization correction (i.e.,
that $N$(Na)$/N$(\NaI) $=10$), a metallicity approximately twice the
solar metallicity, and that
the absorbing gas is at a
distance of $r = 1$ kpc from the nucleus, it is estimated to be flowing
inward at $(1-5)\frac{r}{1~\rm kpc}$ $M_{\odot}~\rm yr^{-1}$.  This
rate is approximately two orders of magnitude larger than the mass
accretion rate required to power a typical Seyfert nucleus 
\citep{crenshaw03,krug10}.  These
cold gas inflows may therefore serve as an important fueling mechanism
for AGN activity in these objects.

Yet further evidence for massive inflows of gas onto AGN-host galaxies
was reported by \cite{stone16}, who studied molecular gas kinematics
in a sample of 52 local AGN selected from the \textit{Swift}-Burst Alert
Telescope Survey of hard X-ray objects.  The \cite{stone16} targets thoroughly
sample the AGN luminosity function to its brightest end. The authors
analyzed spectroscopy of these systems covering the OH $119\mu$m
feature obtained with the \textit{Herschel}/PACS far-IR interferometer,
detecting the transition in absorption in 17 sources.
As in the \cite{krug10} study discussed above, only a handful (four) of
these 17 AGN exhibited molecular outflows, while the OH absorption feature
was redshifted by $>50~\rm km~s^{-1}$ in seven of their targets (corresponding
to a detection rate of $\sim40\%$).  The authors suggested that the
significantly lower detection rates of inflows toward IR-luminous
galaxies may be due to the disruption of these accretion flows by the
faster winds driven by their central starbursts.  %They
%additionally suggest that the high central concentrations of dust in
%IR-bright objects may obscure the nuclear inflow regions from view.  
%I DON'T KNOW ABOUT THAT

%-- Seyferts, so here no issue with NaI stellar absorption profile!
%Inflows in local Volume AGN with Herschel-PACS: Stone et al. 2016

\subsection{Inflows on the Smallest Scales: Feeding Luminous QSOs?}
\label{subsec:qsos}

Spectroscopy of bright quasars
also probes gas flows toward the central source (i.e., the AGN), and can be obtained with much greater efficiency
than faint galaxy spectroscopy.  
Indeed, the SDSS-I/II and SDSS-III BOSS redshift surveys have now revealed
several instances of redshifted
absorption  observed toward bright QSOs \citep{hall02,hall13}.
These profiles occur in quasars with ``broad'' absorption lines (BAL),
defined to
extend over thousands of $~\rm km~s^{-1}$ in velocity \citep{allen11}.  
In addition, the absorption
 typically only partially covers the emitting source.
The vast majority of BAL QSOs exhibit absorption troughs lying
entirely at velocities blueward of systemic, 
and are thought to
arise from AGN-driven feedback/outflow.  
However, in their search through
more than 100,000 SDSS/BOSS QSO spectra, \cite{hall13} discovered 19
BAL objects in which the trough in at least one transition extends to
velocities $v > 3000~\rm km~s^{-1}$.  These authors estimated that such redshifts
occur in approximately 1 in every 1000 BAL quasars.

This work discusses several different physical scenarios which may
explain the observed extreme absorption.  The accretion of material onto the
host dark matter halo or host galaxy at kpc-scale distances from the central
source may certainly contribute to the redshifted absorption signal;
however, the free fall velocity of such material is expected to be only a
few hundred $\rm km~s^{-1}$.  For gas to achieve an infall velocity $v >
3000~\rm km~s^{-1}$, it must reach scales as small as a few hundred
Schwarzschild radii \citep{hall13}.  It remains to be established 
whether infalling gas clumps can maintain sufficiently high densities
at such small distances to give rise to absorption in the observed
transitions (e.g., \SiIV, \CIV, \AlIII\ $\lambda \lambda 1854, 1862$,
\MgII).  Hydrodynamical simulations of accretion flows onto a
supermassive black hole may be used to address this
question; e.g., \cite{li13}; however, 
detailed predictions of absorption profile shapes in multiple
transitions are needed for a quantitative comparison to the
observations.  Alternative scenarios which could give rise to
redshifted, broad absorption include gravitational redshifting of a
spherically symmetric wind; rotating accretion disk winds observed
toward an extended emission source; outflow from a {\it second} QSO
close in the foreground to the first; or even the relativistic Doppler
effect of outflowing ions absorbing photons of lower frequency due to
time dilation.  \cite{hall13} concluded that none of these mechanisms
can individually explain the observations; however, they suggest that
 both rotationally-dominated accretion disk winds and
infalling material are likely contributing to the redshifted broad
absorption profiles.  

In some instances, these data may also constrain the distance between
the absorbing material and the central black hole.  \cite{shi16}
analyzed one of the BAL QSOs discussed in \cite{hall13}, reporting the
detection not only of redshifted \MgII\ and \FeII\ line profiles but also
redshifted hydrogen Balmer and \HeI* $\lambda 3889$ absorption.
The latter transition is highly sensitive to the number of ionizing
photons impinging on the gas per ion ($U$), while Balmer absorption lines
generally arise only in very high-density environments 
\citep[$n$(H) $> 10^6~\rm cm^{-3}$;][]{shi16}.  
Together with an estimate of the ionizing
luminosity of the QSO, constraints on both $U$ and $n$(H) from
analysis of the \HeI* and Balmer lines imply
a distance of $r_{\rm abs} \sim 4$ pc for this particular absorbing system.
\cite{shi16} note that this distance is much larger than expected for
a rotating disk wind or a gravitationally redshifted AGN outflow, and
suggest that it is instead consistent with infall from the inner
surface of a dusty torus surrounding the accretion disk \citep{barvainis87}.
Detailed analysis of a larger sample of similar systems has the potential
to more conclusively establish the frequency of such infall events,
the mass of material involved, and their overall contribution to the
fueling of bright QSOs.  

Apart from the BAL phenomenon, QSOs are observed to exhibit numerous
other classes of foreground absorbers.  One such class -- that of
`proximate' absorbers -- differs from BALs primarily in that they have
significantly more narrow
velocity widths ($< 100~\rm km~s^{-1}$).  Proximate absorbers are
typically defined to have a central velocity within $< 10,000~\rm km~s^{-1}$
of the QSO emission line redshift, and are often (though not always)
observed to fully cover the emitting source \citep{ellison10}.  Recently,
\cite{fath16} presented a sample of six $z\sim2$ QSOs in which a proximate
damped Ly$\alpha$ (DLA) absorber entirely eclipses the broad
Ly$\alpha$ emission from the central AGN.  They report that in five of
these systems, the absorber has a velocity $\sim100-1200~\rm
km~s^{-1}$ redward of the QSO systemic velocity, speculating that the
absorption traces infalling material.  However, because this work
relies on spectroscopy of broad QSO emission lines to constrain the
host galaxy redshift (e.g., \CIV, \CIII]
$\lambda1909$, \HeII\ $\lambda 1640$), these estimates suffer from
significant systematic uncertainties 
\citep[$\sim200-500~\rm km~s^{-1}$;][]{shen16}. 
Followup spectroscopy covering \OII\ $\lambda3728$ or
\OIII\ $\lambda 5007$ in the near-infrared will be important for
verifying this intriguing finding.

\subsection{Inflows onto Early-Type and Post-Starburst Galaxies}

Concomitant with the assembly of this compelling evidence for gas accretion onto
active AGN, two studies offered additional empirical support for
inflow onto red sequence galaxies, or galaxies exhibiting signs of
recent quenching.  
In the first study to obtain rest-frame near-UV galaxy spectroscopy of
sufficient depth for analysis of \FeII\ $\lambda\lambda 2586,2600$ 
and \MgII\ $\lambda\lambda 2796, 2803$ absorption in individual sightlines, 
\cite{coil11}
targeted a sample of 10 X-ray-luminous AGN and 13 post-starburst
galaxies at $0.2 < z < 0.8$ selected from the DEEP2 and SDSS redshift
surveys.  The AGN are moderately luminous, with $\log (L_{X} / \rm
erg~s^{-1}) \sim 41-42$, while the post-starbursts were identified
via a decomposition of their optical spectroscopy into old and young
stellar population components.  Those objects for which spectral
fitting yielded a relatively luminous young component (contributing at
least 25\% of the continuum at $\lambda_{\rm rest} \sim 4500$ \AA) and
which exhibited little or no H$\beta$ emission were targeted.
The Keck/LRIS spectroscopy obtained for this study traced
blueshifted absorption in 60\% of the AGN and 31\% of the
post-starbursts, but also yielded redshifted absorption profiles in two
of the latter.  These redshifts were detected in \MgII\ (and in one case
in \MgI\ $\lambda 2852$) at speeds of $75 - 115~\rm km~s^{-1}$, but were
not detected securely in \FeII\ absorption.  
%These particular objects lie at the brightest, red
%portion of the blue cloud at their respective redshifts
%The SOMETHING TO WRAP UP -- MAYBE THESE PSBs MIGHT FORM MORE STARS?
While it is difficult to draw definitive conclusions from such small
sample sizes, the ubiquitous detection of \MgII\ and \FeII\ absorption in
excess of that predicted for the stellar continuum in
these spectra suggests that the mechanism causing the cessation of
star formation did not completely deplete the gas supply in these systems.

Most recently, \cite{sarzi16} undertook a study of \NaI\ 
absorption in SDSS spectra of
galaxies selected to have high-spatial-resolution 20cm continuum 
coverage obtained with the Very Long Baseline Array \citep{deller14}.  Nearly
$60\%$ of this sample are early-type galaxies, and the vast majority
of these early types do not exhibit blueshifted \NaI.  
The authors briefly comment on their detection of redshifted absorption in
$\sim10-20$ objects, most of which either exhibit very weak optical
emission lines or LINER/Seyfert-like emission line ratios.  It is
suggested that this phenomenon may be the signature of bars or
unsettled dust lanes in these systems.

\subsection{Summary}

In summary, cool gas kinematics have now been assessed in relatively
small samples of early-type and/or AGN-host galaxies.   Measurements
of \NaI\ absorption velocities toward red sequence systems 
suggest a $\sim20\%$ incidence rate of redshifted profiles 
\citep{sato09}.  Spectroscopy of Seyfert galaxies and X-ray-selected
AGN reveals a yet higher incidence ($\sim40\%$) of redshifted \NaI\ or
OH absorption
\citep{krug10,stone16}.  These rates are approximately the same or
higher than the outflow detection rates for the same samples:
$\sim 14\%$ for the red sequence sample of \cite{sato09} and
$\sim10-25\%$ for the Seyfert and BAT/AGN samples of \cite{krug10} and
\cite{stone16}.  While the mass of these flows remains poorly
constrained, the similar rates of these phenomena may ultimately point
to a ``steady state'' of gas cycling through such systems (even as
they maintain very low rates of star formation).  

It should also be
noted, however, that the physical scale over which these flows occur may in fact be
limited to the regions very close to or within  the galaxies' stellar
component.  The common use of \NaI\ in the aforementioned studies naturally
biases these assessments to tracing the motions of cold,
dust-enshrouded material, which in red-sequence galaxies may in fact
lie in dust lanes or cold molecular disks \citep{sato09,davis11}.
Spectroscopy of additional transitions tracing gas over a broader
range of density and temperature (e.g., \MgII\ or \FeII) will be
important for developing constraints on the physical state and scale
of these flows.

%Coil et al. 2011: 2 examples in K+As

%\section{Rest-Frame Ultraviolet Galaxy Spectroscopy: Tracing Inflows Fueling Star-Forming Systems}
\section{Tracing Inflows with Rest-Frame Ultraviolet Galaxy Spectroscopy}
\label{sec:uv}

As discussed above, the initial focus in the literature on \NaI\
absorption kinematics biased these samples toward the highest
surface-brightness systems at $\lambda_{\rm rest} \sim5900$ \AA\ --
i.e., infrared-selected starbursts, massive ellipticals, and bright AGN hosts.  
It was not until large surveys obtaining deep, high-S/N
spectroscopy in the rest-frame ultraviolet were performed that the
first constraints on cool gas kinematics toward a significant ($> 50$)
sample of individual, ``normal'' \textit{star-forming} galaxies were
discussed.  Moreover, the \MgII\ and \FeII\ absorption transitions covered
in these spectra can be significantly stronger than \NaI\ due to a number
of factors; e.g., Mg and Fe are more abundant than Na, and their
singly-ionized transitions trace a much broader range of temperature
and density than neutral Na.  UV galaxy spectra therefore have the potential to trace
more diffuse material, including the halo gas which is known to
give rise to \MgII\ absorption in background QSO sightlines to projected
distances of $R_{\perp} \sim 100$ kpc
\citep{bergeron86,steidel94,kacprzak07,hwc10}. 
Rest-frame UV transitions are thus significantly more sensitive to both inflow
and outflow, and are accessible at
observed-frame optical wavelengths for galaxies at $z>0.3$.

\begin{figure}[t]
% For example, with the graphicx style use
\hskip -1cm
\includegraphics[angle=90,scale=0.4,viewport=0 0 400 470,clip]{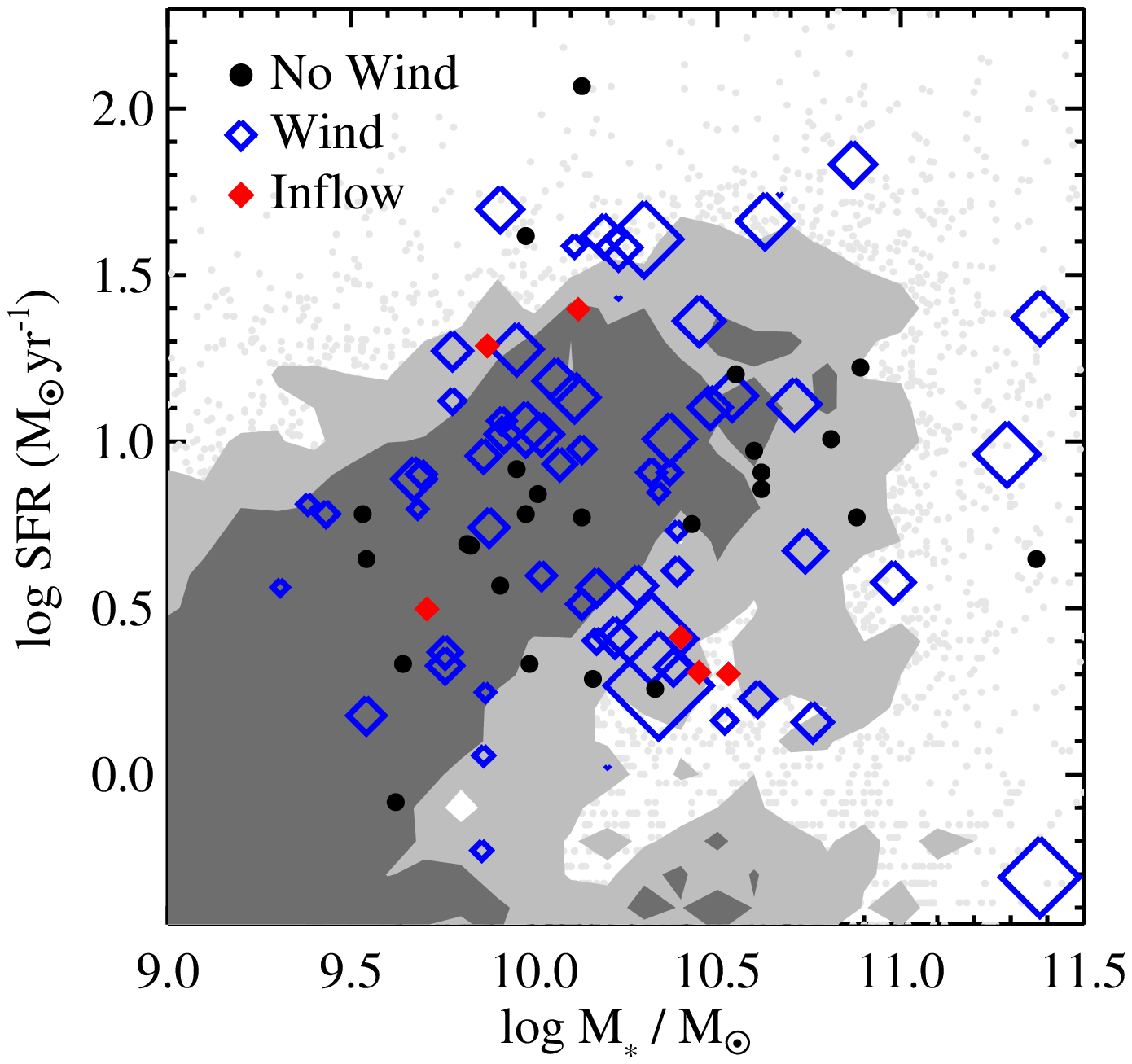}
\hskip -1cm
\includegraphics[angle=90,scale=0.4,viewport=0 0 400 470,clip]{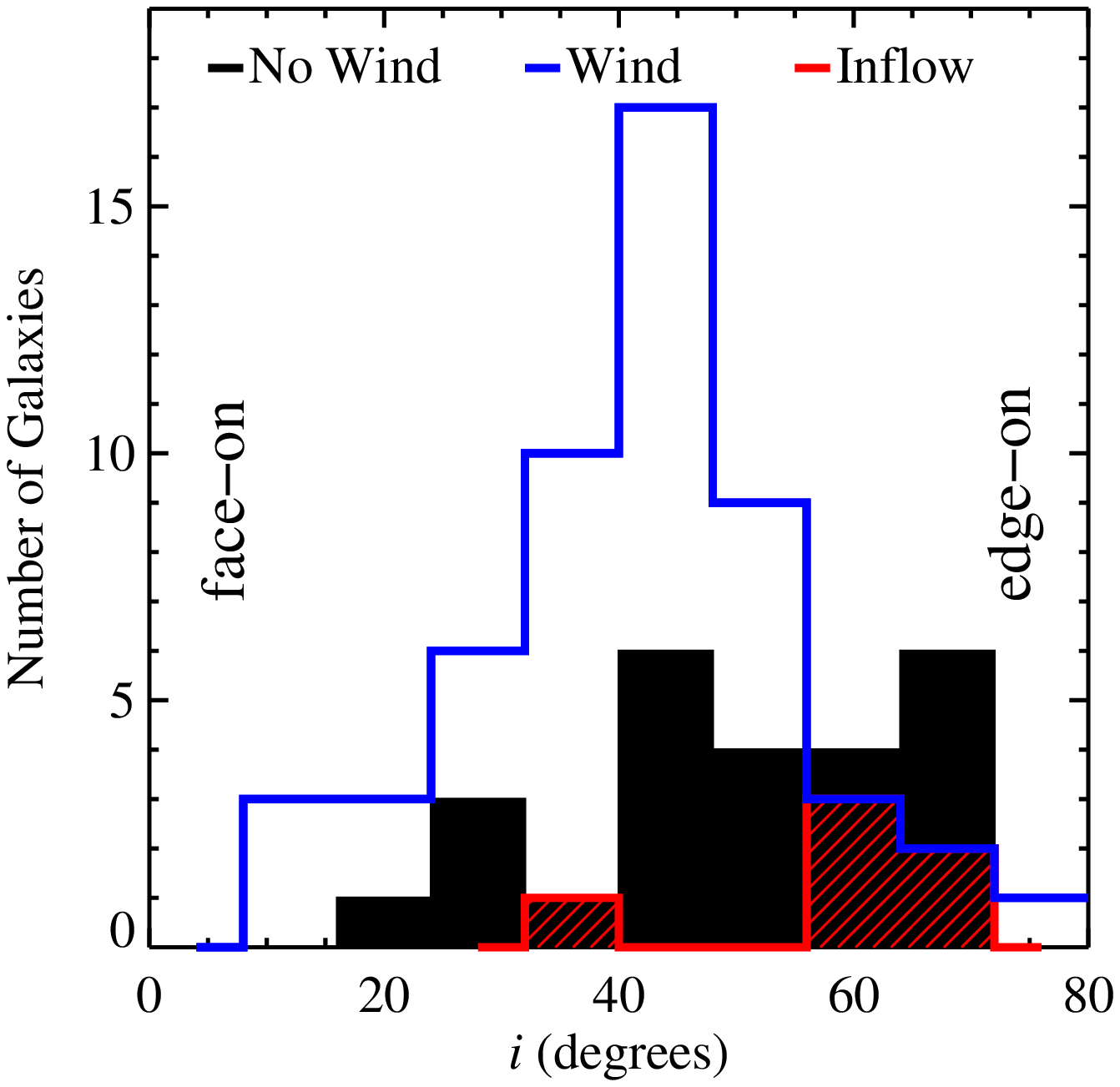}
%\includegraphics[angle=90,scale=0.4,viewport=0 0 400 470,clip]{incl_hist_resprop.pdf}
%
% If no graphics program available, insert a blank space i.e. use
%\picplace{5cm}{2cm} % Give the correct figure height and width in cm
%
\caption{\textbf{The first survey to detect cool gas inflow traced by \MgII\
  and \FeII\ absorption onto a sample of distant star-forming galaxies
  \citep{rubin12,rubin14}}.  \textit{Left:}  Colored and black points
show the sample of $\sim100$ galaxies targeted in
\cite{rubin12,rubin14}, and gray contours show the SFR-$M_*$
distribution of the underlying galaxy population \citep{barro11}.  Blue
open diamonds indicate galaxies with detected winds, red filled
diamonds mark objects with inflows, and black circles indicate objects
with neither winds nor inflow.  The size of the blue diamonds
is scaled with the \MgII\ equivalent width of the outflow.  
Winds are detected in $\sim2/3$ of the
galaxies, while detected inflows occur at a rate $\sim6\%$.
\textit{Right:} Distribution of inclinations ($i$) for the disk-like
galaxies selected from the same study.  The distribution for galaxies
with winds is shown in blue, and the distributions for galaxies with
inflows and without winds/inflows are shown in red and black,
respectively.  While face-on galaxies  (at low $i$) are significantly
more likely to drive a detected wind than galaxies viewed edge-on,
the galaxies with detected inflow are nearly all highly inclined.
Panels are adapted from \cite{rubin12,rubin14}.}
\label{fig:2}       % Give a unique label
\end{figure}

The first unambiguous detection of inflow observed down the barrel
toward a sample of star-forming galaxies was reported in
\cite{rubin12}.  The galaxies were targeted over the course of a
high-S/N Keck/LRIS survey of $\sim100$ galaxies at redshifts
$0.3<z<1.4$ \citep{rubin14}.  This sample, selected to a magnitude
limit $B_{\rm AB} < 23$, spans the star-forming sequence at $z\sim0.5$, and is
thus representative of the ``normal'' star-forming galaxy population
(Fig.~\ref{fig:2}, left).
In addition, this survey targeted fields with deep
\textit{HST}/ACS imaging, facilitating a detailed morphological
analysis (Fig.~\ref{fig:2}, right).

Blueshifted absorption tracing outflows was detected in the majority
($\sim66\%$) of this sample.  Moreover, the detection rate of these
winds does not vary significantly with the star formation rate (SFR) or stellar mass of the
host, but rather depends primarily on galaxy orientation
(Fig.~\ref{fig:2}, right).  This finding is suggestive not only of
ubiquitous outflows, but also of an approximately biconical morphology
for these flows across the star-forming sequence.  

\begin{figure}[]
%\sidecaption
\includegraphics[angle=90,scale=0.6,viewport=0 785 320 330,clip]{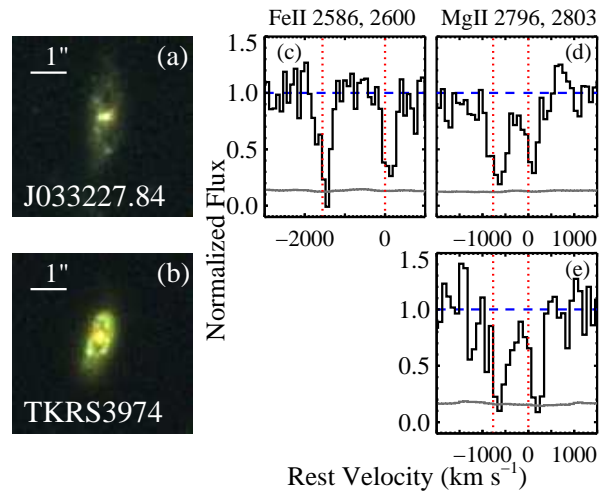}
\caption{\textbf{Imaging and spectroscopy of two
    galaxies with ongoing gas inflow.}  The left column shows
  \textit{HST}/ACS images, and the middle and right columns show \FeII\
  and \MgII\ transitions in the galaxy spectra.  The line profiles are
  redshifted with respect to systemic velocity (marked with vertical
  dotted lines).  Analysis of the imaging indicates that these
  galaxies are disk-like with edge-on orientations.
The figure is a reproduction of a portion of Figure 1 from the article
``The Direct Detection of Cool, Metal-Enriched Gas Accretion onto
Galaxies at $z\sim0.5$'', by K. H. R. Rubin et al.\ (2012, ApJL, 747,
26).  \textcopyright AAS.  Reproduced with permission.}
\label{fig:3}       % Give a unique label
\end{figure}

In the same survey, redshifted absorption tracing cool inflow was detected in six of the
remaining galaxy spectra (red diamonds in Fig.~\ref{fig:2};
Fig.~\ref{fig:3}) with velocities  $\sim 80-200~\rm km~s^{-1}$.
The galaxies themselves have SFRs ranging from $\sim1-40~M_{\odot}~\rm
yr^{-1}$, and have stellar masses in the range $9.6 < \log
M_*/M_{\odot} < 10.5$.  Perhaps most significantly, five of these six
galaxies have disk-like morphologies and are viewed in a nearly
edge-on orientation (with inclinations $> 55^{\circ}$;
Fig.~\ref{fig:2}, right).  The authors suggest that the preferential
detection of inflows toward edge-on galaxies indicates that cool
infall is more likely to occur along the plane of galactic disks,
rather than along the minor axis.
We also note here that higher sensitivity to inflows in more edge-on systems is a
natural consequence of biconical winds.  

Assuming a metallicity for these flows of $Z = 0.1 Z_{\odot}$ and
adopting constraints on the inflow \MgII\ and \FeII\ column densities from
their absorption line analysis, \cite{rubin12} estimated 
lower limits for the mass accretion rates onto this sample of six
objects in the range  $\sim0.2-3~M_{\odot}~\rm yr^{-1}$.  These observed flow rates are approximately consistent
with the rate of mass flow onto our own Galaxy \citep{lh11}.

\begin{figure}[]
\sidecaption
% For example, with the graphicx style use
\includegraphics[scale=4.7,trim=-8 0 -8 0,clip]{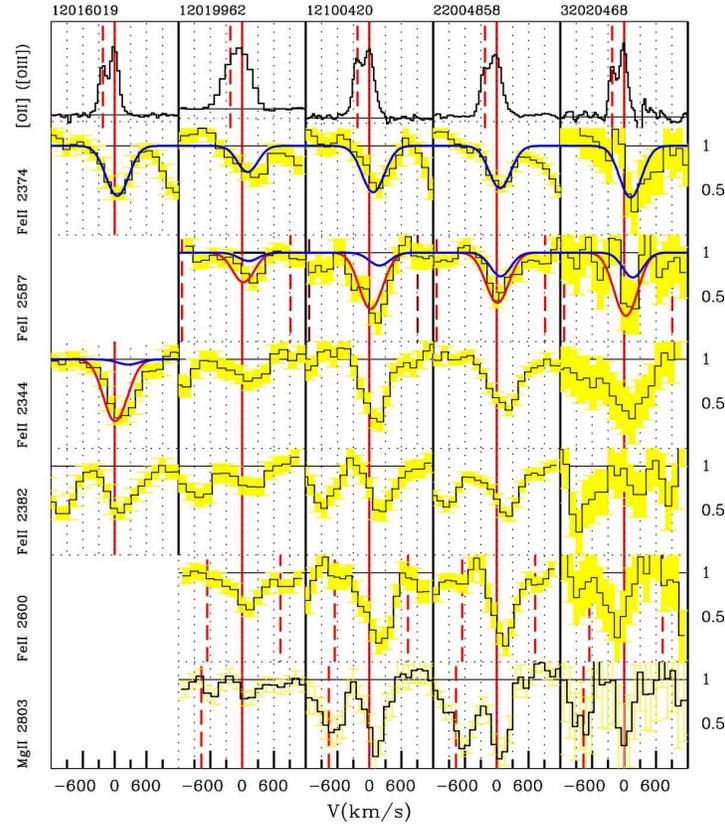}
%
% If no graphics program available, insert a blank space i.e. use
%\picplace{5cm}{2cm} % Give the correct figure height and width in cm
%
\caption{\textbf{Resonant absorption lines and \OII\ emission profiles
  for five galaxies with securely-detected inflows from
  \cite{martin12}.}  
Galaxy ID numbers are indicated at the top of each column.
The solid red vertical lines mark the rest-frame velocity of each
transition, and dashed vertical lines mark the wavelengths of nearby resonant
transitions.  Single-component absorption line fits are shown in the
row of \FeII\ $\lambda2374$ profiles, and fits including both an
absorption component with a velocity fixed at $0~\rm km~s^{-1}$ and a
``flow'' component are overplotted on the \FeII\ $\lambda 2586$
profiles.  These objects exhibit strong redshifts in most of the
absorption transitions shown.  The \OII\ profiles are used to
determine the systemic velocity.
This figure is a reproduction of Figure 16 from the article 
``Demographics and Physical Properties of Gas Outflows/Inflows at $0.4
< z < 1.4$'', by C. L. Martin et al.\ (2012, ApJ, 760, 127).  \textcopyright AAS. Reproduced with permission.}
\label{fig:4}       % Give a unique label
\end{figure}

Nearly concurrently, \cite{martin12} carried out a similar but
completely independent study of
cool gas kinematics in Keck/LRIS spectroscopy of $\sim200$ galaxies.  
The galaxy sample has a somewhat higher median redshift than that of
\cite{rubin12} ($<z> \sim 0.5$ vs.\ $\sim1$) and is selected to a fainter
magnitude limit $B_{\rm AB} < 24.0$.  However, the vast majority of
the targets are star-forming, with SFRs ranging from
$\sim1-98~M_{\odot}~\rm yr^{-1}$ and stellar masses $8.85 < \log
M_*/M_{\odot} < 11.3$.  This survey made use of a somewhat
lower-resolution spectroscopic setup for $\sim70\%$ of the sample
(having a FWHM resolution element $\sim435~\rm km~s^{-1}$
vs.\ $\sim282~\rm km~s^{-1}$), which in principle limits sensitivity
to lower-velocity flows.  In spite of this, redshifted absorption
profiles were detected in nine spectra (see Fig.~\ref{fig:4}), yielding a detection rate ($\sim4\%$)
 consistent with that of \cite{rubin12}.  \cite{martin12} also
performed a careful analysis of the two-dimensional spectra of these
objects, identifying weak nebular line emission offset from the
continuum trace at the same velocity as the redshifted absorption in a
few (4) cases.  They speculated that the inflows in these
systems are
being fed by relatively dense, star-forming structures (e.g., satellite dwarf galaxies) rather than diffuse
accretion streams from the circumgalactic medium.  

\cite{martin12} found that the galaxies exhibiting inflows span the
range in stellar mass and SFR occupied by the parent sample.  They
also noted that among the four inflow galaxies for which quantitative
morphologies are available, only one has a high inclination
($i\sim61^{\circ}$); the remaining three galaxies have
$i<55^{\circ}$ \citep{kornei12}.  The detection of inflow toward these
objects requires that the idea put forth by \cite{rubin12}
that infall is more likely to be observed along the plane of galactic
disks be considered more carefully and investigated with a significantly
larger spectroscopic sample.

These two studies have provided us with the first, unequivocal
evidence for gas accretion onto distant, star-forming galaxies.
However, the physical nature of these flows remains an open
question.  There are numerous potential sources for the enriched
material producing the observed absorption, including gas which has
been tidally stripped from nearby dwarf galaxies, or wind ejecta from
the central galaxy which is being recycled back to the disk.  
Indeed, such wind recycling is predicted in numerous cosmological
galaxy formation simulations \citep{opp10,vogelsberger13}.  Current
absorption line data cannot distinguish between these alternatives;
however, detailed constraints on the covering fraction or cross section of the
inflow from upcoming integral field spectroscopic surveys may aid in
differentiating between these scenarios.  This topic will be discussed
further in Sections~\ref{subsec:ifus} and \ref{sec:future}.

%%Giavalisco?? -- do I have to??

%\subsubsection{Subsubsection Heading}
%\footnote{If you copy
%text passages, figures, or tables from other works, you must obtain
%\textit{permission} from the copyright holder (usually the original
%publisher). Please enclose the signed permission with the manuscript. The
%sources\index{permission to print} must be acknowledged either in the
%captions, as footnotes or in a separate section of the book.}

%\paragraph{Paragraph Heading} %

\section{Toward Assessment of the Incidence of Inflow}
\label{sec:biases}
% Always give a unique label
% and use \ref{<label>} for cross-references
% and \cite{<label>} for bibliographic references
% use \sectionmark{}
% to alter or adjust the section heading in the running head

%We must also consider scenarios in which our absorption line technique
%will be insensitive to the signature of gas inflow along the line of
%sight.  
Ultimately, empirical studies characterizing the cool gas flows around
galaxies must constrain, e.g., the
incidence of accretion in a given galaxy population as a function of
the mass inflow rate and age of the universe.  Such detailed
assessments are required for an incisive test of galaxy formation models 
 \citep{dave12,nelson15}.  We now consider a
few factors which complicate the use of current datasets in developing
such constraints.  In the following section we discuss future directions which may ameliorate these issues.

\subsection{Spectral Confusion}
\label{subsec:resolution}

The faintness of the galaxies studied in \cite{rubin12} and
\cite{martin12} forced the selection of a low resolution
spectroscopic setup for these surveys (FWHM $\sim 250-400~\rm
km~s^{-1}$).  Thus, the velocity profiles arising from distinct inflowing
and outflowing gas structures along the same line of sight are blended in these spectra.  Indeed, in none of the 
\MgII\ or \FeII\ line profiles discussed in \cite{rubin14} is more than one velocity component
resolved.  If there is, e.g., a component of gas being accreted at
$v\sim+100~\rm km~s^{-1}$, and another wind component foreground to the same
galaxy beam moving at $v\sim-200~\rm km~s^{-1}$, a centroid of the resulting
line profile (observed at low resolution) will suggest the presence of an outflow with a velocity
$v\sim-100~\rm km~s^{-1}$ (while the inflowing component is completely
obscured).  Or, in the case of two components each at $v\sim+100~\rm
km~s^{-1}$ and $-100~\rm km~s^{-1}$, a centroid of the observed line
profile will reveal neither outflow nor inflow.  
Thus, there is a limited set of circumstances in
which the centroid of such blended profiles will be significantly redshifted,
revealing an inflow.  

\begin{figure}[]
%\sidecaption
% For example, with the graphicx style use
\includegraphics[angle=90,scale=0.4]{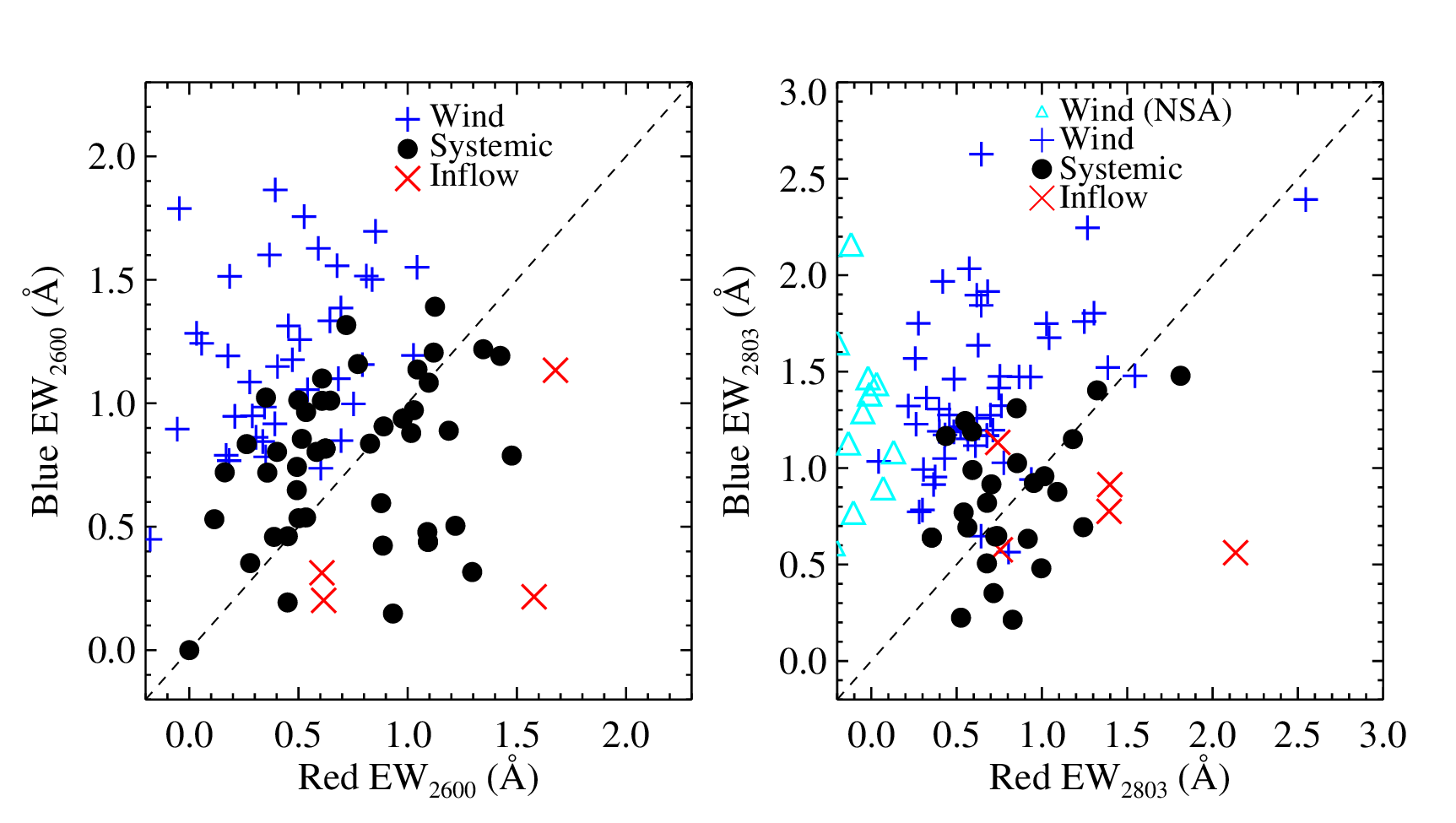}
%
% If no graphics program available, insert a blank space i.e. use
%\picplace{5cm}{2cm} % Give the correct figure height and width in cm
%
\caption{\textbf{Assessment of the symmetry of \FeII\ and \MgII\ absorption line profiles in
  low-resolution galaxy spectroscopy.}  
All measurements shown are from analysis of the UV spectroscopy of
$\sim100$ star-forming galaxies discussed 
 in \cite{rubin12,rubin14}.
\textit{Left:}  The equivalent width (EW) of the \FeII\ $\lambda 2600$
transition, measured at $v<0~\rm km~s^{-1}$ (``blue''), vs.\ the EW of
the same transition at $v>0~\rm km~s^{-1}$ (``red'') for galaxies with
blueshifted \FeII\ (blue crosses), without winds or inflows (black
filled circles), and with inflows detected in \FeII\ (red crosses).
Several galaxies without detected inflows have red $\rm EW_{2600}$
values $> 0.6$ \AA, comparable to or larger than the red $\rm
EW_{2600}$ for two of the inflow galaxies.  
\textit{Right:} same as left-hand panel, for the \MgII\ $\lambda 2803$
transition.  Spectra which have no detectable absorption at $v>0~\rm
km~s^{-1}$ are marked with cyan triangles.  As is the case for \FeII\
$\lambda 2600$, numerous galaxies without detected inflows have red
$\rm EW_{2803} > 0.8$ \AA, the smallest value measured for the inflow
sample. 
This survey is therefore likely failing to flag ongoing inflow in a
significant fraction of those galaxies in the ``wind'' and
``systemic'' subsamples.
This Figure is a reproduction of Figure 7 from the article 
``Evidence for Ubiquitous Collimated Galactic-Scale Outflows Along the
Star-Forming Sequence at $z\sim0.5$'', by K. H. R. Rubin et al.\
(2014, ApJ, 794, 156).  \textcopyright AAS. Reproduced with permission.}
\label{fig:5}       % Give a unique label
\end{figure}

Indeed, as noted in \cite{rubin12}, the six objects toward which
inflows were detected are unique not for the material moving at
positive velocities along the line of sight, but for the
\textit{absence of winds}.  This is demonstrated in Fig.~\ref{fig:5},
which compares the equivalent widths measured redward and blueward of
systemic velocity in the \FeII\ $\lambda2600$ (left) and \MgII\ $\lambda2803$ (right)
transitions in the spectral sample of \cite{rubin14}.  Objects with
detected winds are marked with blue crosses and cyan triangles; objects with detected
inflows are marked with red crosses; and objects with neither winds
nor inflow are marked with black filled circles.  A 1:1 relation is
indicated with the dashed lines.  Inflows were detected in galaxies with
larger red EWs than blue EWs; however, there are many other
systems which have red EWs larger than the smallest values measured for
the inflow sample (i.e., red $\rm EW_{2600} > 0.6~\AA$ and red $\rm
EW_{2803} > 0.8~\AA$; see Fig.~\ref{fig:5}).  Based on an accounting of the number of
galaxies with red EWs at least as large as those in which inflow was
detected, \cite{rubin12} estimated that inflow must be occurring in at least
20-40\% of the full galaxy sample.  

%This systematic issue is certainly reduced at higher spectral
%resolution.   
Sensitivity to inflow in the presence of absorption due to winds or
interstellar material is certainly improved at higher spectral resolution.
The \NaI\ spectroscopy of nearby Seyfert galaxies discussed in \cite{krug10}
(see Section~\ref{subsec:agn}) has a FWHM velocity resolution of
$\sim85~\rm km~s^{-1}$, and is thus significantly more sensitive to
detailed line profile shapes than that of \cite{rubin12} or
\cite{martin12}.  %Even so, the spectra shown in Fig.\ 1 of
%\cite{krug10} exhibit strongly blended velocity components.  
Such high
spectral resolution can be achieved with current technology for distant 
galaxies which are also strongly gravitationally lensed
\citep[e.g.][]{pettini02,quider09,quider10,dz10}.  If at $z>2$,
high-resolution optical spectroscopy of these
targets offers the added advantage of coverage of a rich suite of
rest-frame UV transitions, including Ly$\alpha$, \OI\ $\lambda 1302$,
several \SiII\ transitions, \SiIV\ $\lambda \lambda 1393, 1402$, and \CIV\
$\lambda \lambda 1548, 1550$.

%\subsection{Subsection Heading} %

%\begin{svgraybox}
%If you want to emphasize complete paragraphs of texts we recommend to use the newly defined Springer class option \verb|graybox| and the newly defined environment \verb|svgraybox|. This will produce a 15 percent screened box 'behind' your text.

%If you want to emphasize complete paragraphs of texts we recommend to use the newly defined Springer class option and environment \verb|svgraybox|. This will produce a 15 percent screened box 'behind' your text.
%\end{svgraybox}

\begin{figure}[]
\sidecaption
\includegraphics[angle=90,scale=0.75,viewport=0 0 400
270,clip]{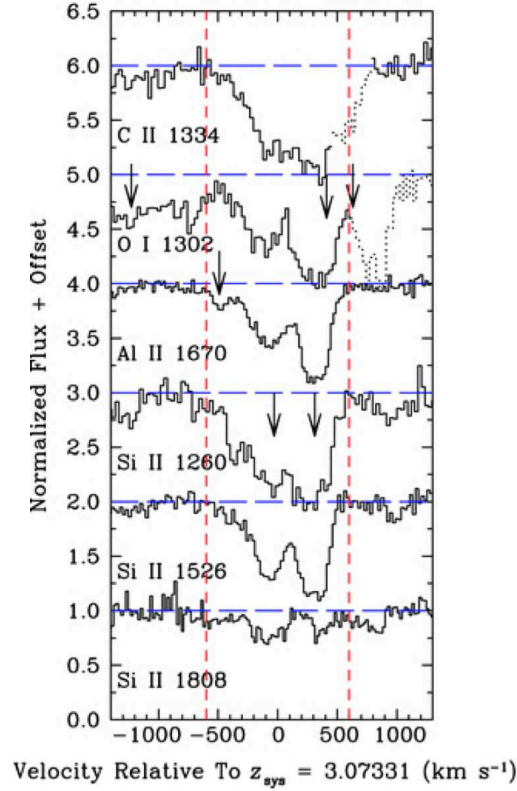}
%\raisebox{5.3in}{
%  \includegraphics[angle=270,scale=0.6,viewport=0 -20 600
%  280,clip]{quider10_f4.eps}
%}
%
% If no graphics program available, insert a blank space i.e. use
%\picplace{5cm}{2cm} % Give the correct figure height and width in cm
%
\caption{\textbf{High resolution Keck/ESI spectrum showing rest-frame UV absorption
    lines in a lensed Lyman break galaxy at $z_{\rm sys}=3.073$} published in
  \cite{quider10}.  Arrows mark absorbers unrelated to this system.
  The values on the x-axis indicate velocity relative to $z_{\rm sys}$.
  The absorption transitions shown trace either cool, photoionized
  material (e.g., \CII, \AlII, \SiII) or neutral gas (\OI) along the
  sightline.  The spectrum resolves two absorbing ``components'', one
  at $v\sim -70~\rm km~s^{-1}$ and one at $v\sim+350~\rm km~s^{-1}$,
  suggesting both ongoing outflow and accretion.  %\cite{quider10}
  %noted 
  %that if this object were observed at low resolution ($R<1000$), such
  %that these components are blended, the absorption profiles would
  %have a central velocity of $\sim +150~\rm km~s^{-1}$.  
Spectroscopy of similar quality and fidelity of a significantly larger
galaxy sample will reveal the frequency with which these phenomena
occur along the same line of sight.
This Figure
  reproduces a portion of Figure 4 from the article ``A study of
  interstellar gas and stars in the gravitationally lensed galaxy `the
  Cosmic Eye' from rest-frame ultraviolet spectroscopy'' by
  A. M. Quider et al.\ (2010, MNRAS, 402, 1467).  Reproduced with the
  permission of Oxford University Press.}
\label{fig:6}       % Give a unique label
\end{figure}

Keck/ESI coverage of many of these transitions in a spectrum of the 
``Cosmic Eye'' is shown in Fig.~\ref{fig:6},
adapted from \cite{quider10}.  The Cosmic Eye is a Lyman break galaxy (LBG)
at $z=3.073$ which is heavily magnified by foreground massive
structures, producing two overlapping arcs on the sky.
The Keck/ESI spectroscopy obtained by \cite{quider10} has a FWHM
velocity resolution $\sim75~\rm km~s^{-1}$, and successfully resolves at
least two distinct absorption components in several transitions.  In
fact, the authors note that this spectrum exhibits a strong,
redshifted absorption component with $v\sim350~\rm km~s^{-1}$ in
addition to a blueshifted component presumably arising from winds.
This may be interpreted as yet another detection of gas accretion, the
first at $z\sim3$.  Indeed, the red absorption component in this line
profile was unique among the lensed LBGs that had been studied to
date, each of which exhibited strongly (and exclusively) blueshifted absorption.

Moreover, these data may in principle be used to constrain the column
density, metallicity, and mass of the absorbing gas to a substantially
higher level of precision than is possible with spectroscopy covering
only the \NaI\ or \MgII\ and \FeII\ lines.
The absorption in the latter two transitions is nearly always saturated, 
limiting the line-of-sight column density to be larger than a
modest value
\citep[e.g., $N_{\rm MgII}\sim10^{14}~\rm cm^{-2}$ or 
$N_{\rm H}\sim10^{18.4}~\rm cm^{-2}$ assuming solar 
metallicity;][]{rubin12,martin12}. 
Furthermore, rest-frame optical and near-UV spectroscopy cannot be used to 
measure
the column density of hydrogen toward the galaxy, nor does it cover
transitions of a given element in multiple levels of ionization for
constraints on the ionization state of the material.
%there are numerous, weaker ionic metal transitions in the far-UV which 
While the S/N of the
spectrum of the Cosmic Eye discussed above is insufficient for such analysis,
spectroscopy of other lensed LBGs, e.g. the ``Cosmic Horseshoe''
\citep{quider09}, MS1512-cB58 \citep{pettini02}, and the ``8 o'clock arc''
\citep{dz10} has yielded ionic column density measurements for ISM
and outflow absorbing components to a
precision of $\pm0.1$ dex via analysis
of weak (and unsaturated) \SiII\ and \FeII\ transitions.  Combined
with analysis of the Ly$\alpha$ line profiles in the same spectra, 
this work has offered some of the only constraints on the
metallicity of gas known to be outflowing from
distant galaxies.  Deeper observations 
of the Cosmic Eye and expanded samples of high dispersion, rest-frame
far-UV spectroscopy of lensed LBGs have the potential to yield
important constraints on the mass and metallicity of both outflows and
cool galactic gas accretion.

\subsection{Spatial Resolution}
\label{subsec:ifus}

\begin{figure}[]
%\sidecaption
% For example, with the graphicx style use
\includegraphics[scale=0.4]{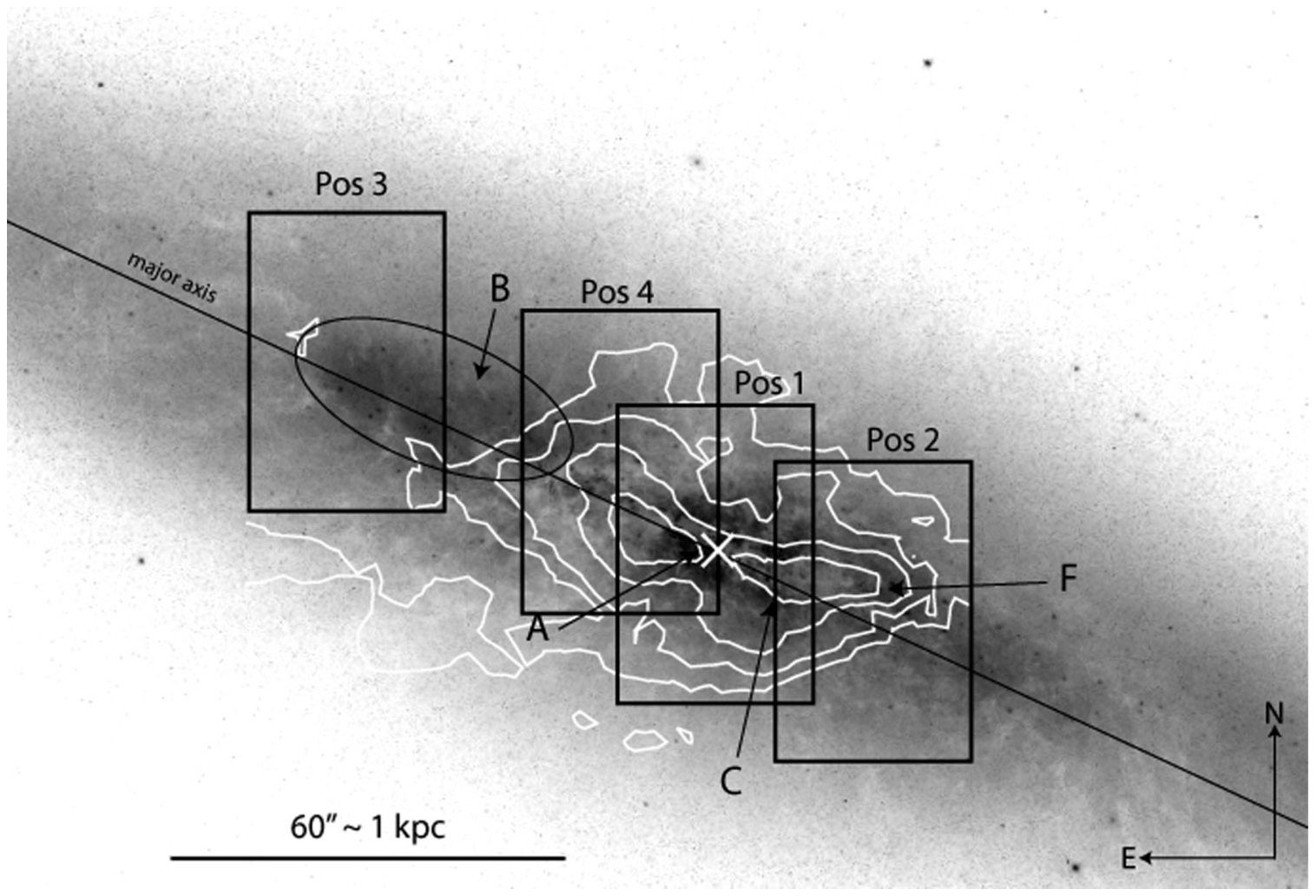}
\includegraphics[scale=0.8,viewport=0 0 167 200,clip]{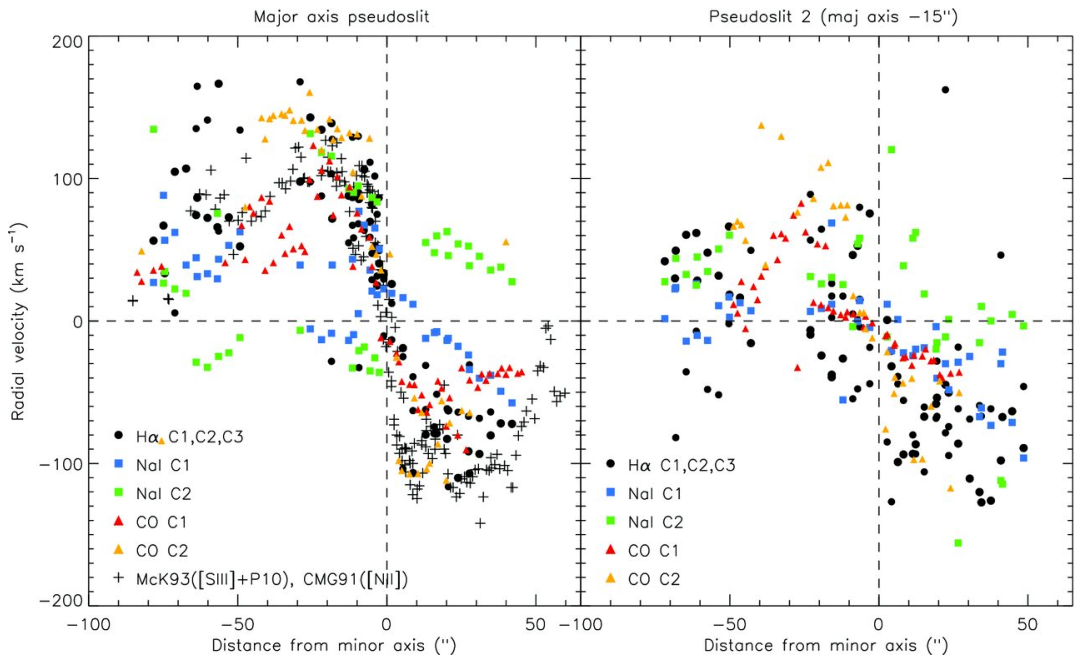}
%
% If no graphics program available, insert a blank space i.e. use
%\picplace{5cm}{2cm} % Give the correct figure height and width in cm
%
\caption{\textbf{Spatially resolved spectroscopy of the starburst
    galaxy M82.}  \textit{Left:} \textit{HST}/ACS image of M82 with
  four IFU pointings overlaid (rectangles).  The white contours show
  the total CO flux from \cite{walter02}.
\textit{Right:} Radial velocity of \NaI, H$\alpha$, and CO measured in
a 7 arcsec-wide pseudo-slit extracted from the IFU data along the major axis.  
 The CO and H$\alpha$
velocities are similar across the galaxy; however, there is an absorbing
component of \NaI\ (green) which is offset in velocity by many tens of $\rm km~s^{-1}$.
\cite{westmo13} suggest that this
component may be due to accreting tidal debris.
Panels are reproductions of Figures 1 (left) and 7 (right) from the
article ``Spatially resolved kinematics of the multi-phase
interstellar medium in the inner disc of M82'' by M. S. Westmoquette
et al.\ (2013, MNRAS, 428, 1743).  Reproduced with the
  permission of Oxford University Press.}
\label{fig:7}       % Give a unique label
\end{figure}

In addition to increasing spectral resolution to gain sensitivity to
gas accretion, spatially resolving the background stellar beams will
further improve our constraints on the morphology of inflow.  
\cite{westmo13} offered 
one of the earliest demonstrations of the potential of this technique, 
targeting the central regions of
the M82 starburst with four pointings of the DensePak IFU on the WIYN
telescope (Fig.~\ref{fig:7}).  These observations achieved 3-arcsec
spatial resolution at a spectral resolution of $\sim45~\rm
km~s^{-1}$.  Combined with CO emission observations from \cite{walter02},
they facilitated a detailed comparison between the
kinematics of H$\alpha$ emission, CO emission, and \NaI\ absorption.

The authors extracted spectra from a pseudo-slit 7 arcseconds in width 
placed along the major axis of the system, co-adding the data in
several spatial bins along the slit.  Measurements of the H$\alpha$, \NaI, and
CO velocities in these bins are shown in the right-hand panel of
Fig.~\ref{fig:7}.
In general, these velocities are similar at all slit locations;
however, there is one component of \NaI\ absorption (green) which has
a large velocity offset, appearing to
counter-rotate.  The authors suggest that this component could be due
to infalling tidal debris, which may even have originated in the HI
gas filaments populating the circumgalactic medium around this system.  

While this interpretation is somewhat speculative, this spatially
resolved spectroscopy has nevertheless resolved a heretofore unknown
velocity component of cold gas in a system in which gas flows have
been studied in great detail for over a decade \citep{shopbell98}.
Moreover, these data provide evidence in support of a picture in which
inflows persist even in strongly starbursting systems with powerful ongoing outflows.
Upcoming massive IFU
surveys such as SDSS-IV/MaNGA \citep{bundy15} will provide
qualitatively similar observations for $\sim10,000$ nearby galaxies,
resolving \NaI\ kinematics on $\sim1-2$ kpc spatial scales
across the face of each object.  These data
will be extremely sensitive to the morphology and cross
section of inflowing streams as they approach the ISM of galaxies.

\section{Summary and Future Directions}
\label{sec:future}

Deep galaxy spectroscopy has now provided perhaps the only
unequivocal evidence for the inflow of gas toward galaxies beyond the
local universe.   While the earliest spectroscopic datasets
probing cool gas kinematics ``down the barrel'' specifically targeted starbursting
galaxies and typically reported only blueshifted (i.e., outflowing)
absorption \citep{heckman00,rupke05b,martin05}, later surveys achieving the requisite S/N have revealed
numerous instances of redshifted metal-line absorption profiles toward a wide
variety of host systems. Study of \NaI\ D profiles in spectra of
red-sequence galaxies at $z\sim0.3$
 has suggested an incidence rate of inflow of $\sim20\%$;
i.e., similar to the rate of outflow detections for the same sample
\citep{sato09}.  Spectroscopy of \NaI\ and the OH $119\mu$m feature in
the far-IR in small samples of
nearby Seyfert and
X-ray-bright AGN hosts points to a yet higher rate of redshifted cool
gas absorption \citep[$\sim40\%$;][]{krug10,stone16}.  
Such inflows may in fact be necessary to fuel the observed nuclear
activity; however, the amount of mass carried in these flows remains
poorly constrained, and current observations do not establish the
present (or ultimate) location of the gas along the line of sight
\citep[although see][for an exception to this generalization]{shi16}.  

In the few years following these first detections, deep spectroscopy
of \FeII\ and \MgII\ transitions 
in the rest-frame near-UV has finally revealed evidence for inflow
onto ``normal'' star-forming galaxies at $z > 0.3$ \citep{rubin12,martin12}.
The reported rate of incidence is low ($<10\%$), as blueshifted
absorption tends to dominate the metal-line profiles in these
spectra.  Furthermore, \cite{rubin12} has presented evidence suggesting that inflow is
more likely to be detected in these star-forming systems when they are viewed in an edge-on
orientation.  However, \cite{martin12} reported detections of  inflow toward galaxies
viewed over a wide range of orientations, and hence do not support
this claimed dependence of inflow detection rate on viewing angle.

Such studies have been pivotal for furthering our understanding of the
cycling of cool gas through galaxy environments.  However, the evidence
they offer is anecdotal rather than statistical.  A complete,
empirical picture of the baryon cycle must establish the incidence,
mass, and morphology of gas inflow as a function of host galaxy
stellar mass, star formation activity and history, and AGN
luminosity.  
As all of the
aforementioned samples are S/N-limited, they cannot 
assess such quantities regardless of their spectral resolution or the
transitions they probe.  Surveys of cool gas kinematics in
samples selected to be complete to a given stellar mass limit (i.e., unbiased
in their distributions of SFR and galaxy orientation) are required 
if we are to make substantive progress in our development of an empirical model of galactic gas
flows.

At the same time, the observational challenges inherent in such an
effort are significant.  The works discussed in Section~\ref{sec:uv} surveying samples of $\sim100-200$
objects in the rest-frame near-UV together represent an investment of  $\sim17$
nights on the Keck 1 Telescope \citep{rubin12,martin12}.
A stellar mass-complete survey of a significantly larger sample would
require a yet more extensive observing campaign. The ongoing
SDSS-IV/MaNGA survey \citep{bundy15} will facilitate a major  
advancement in our constraints on the incidence and morphology of inflow
in the near term, as it will obtain high-S/N, spatially resolved
spectroscopy of \NaI\ in an unprecedented sample of $\sim10,000$
nearby galaxies.  The MaNGA sample selection is carefully designed to be
complete to a stellar mass $\log M_*/M_{\odot} > 9$, and its observing
procedure ensures a minimum S/N per fiber of $\sim6$ per pix at $1.5$
effective radii.  Although its spectral coverage is limited to \CaII\
and \NaI\ at relatively low velocity resolution ($R\sim2000$), its
$\sim1-2$ kpc spatial resolution will enable detailed mapping of the
incidence of cold, dusty
gas inflow at speeds $> 40~\rm km~s^{-1}$ in an extremely large galaxy
sample.  

%In the near term, MaNGA will reveal inflow in both SF and red
%galaxies, but we need to make sure we understand the stellar continuum
%around NaI

To assess the mass and metallicity of these flows, deep, high
resolution rest-frame far-UV spectroscopy 
 will be required \citep{quider09,quider10,dz10}.
Such observations of samples of more than a few objects must await
the next generation of wide-field multi-object spectrographs on 
30m-class ground-based optical telescopes.  As laid out in the Thirty
Meter Telescope
Detailed Science Case \citep{tmt}, the prospective instrument WFOS
will be capable of obtaining $R\sim5000$ spectroscopy of more than $100$ 
(unlensed) LBGs at $z\sim2-3$  with $R_{\rm AB} < 24.5$ simultaneously.  
S/N $> 30$ will be achieved in just a few hours for these spectra,
which will cover all of the ionic transitions discussed in the context
of the Cosmic Eye  (Section~\ref{subsec:resolution}),
including Ly$\alpha$, \OI,
several \SiII\ transitions, several \FeII\ transitions, \AlIII, \SiIV, and \CIV.
Such a dataset will
permit detailed constraints on the column densities, ionization
states, metallicities, and mass of gas components arising in the ISM,
outflows, and in accreting streams at high redshift.

Similarly detailed characterization of gas flows at $z<1.5$ must await
the next UV-sensitive space mission.  A prospective high-resolution
imaging spectrograph \citep{france16} conceived for the Large
Ultraviolet/Optical/InfraRed (LUVOIR) surveyor NASA mission concept
will not only access the important UV transitions discussed above, but
will do so at a spatial resolution of $\sim10-100$ pc.  This
instrument will readily differentiate between material flowing inward
in accreting streams from ongoing outflow and
establish the mass and morphology of this accreting material.  
These capabilities will 
ultimately allow us to complete our empirical picture of the cycling of diffuse
baryons through galaxy environments.

%\begin{acknowledgement}
%If you want to include acknowledgments of assistance and the like at the end of an individual chapter please use the \verb|acknowledgement| environment -- it will automatically render Springer's preferred layout.
%\end{acknowledgement}
%
%\section*{Appendix}
%\addcontentsline{toc}{section}{Appendix}
%
%
%When placed at the end of a chapter or contribution (as opposed to at the end of the book), the numbering of tables, figures, and equations in the appendix section continues on from that in the main text. Hence please \textit{do not} use the \verb|appendix| command when writing an appendix at the end of your chapter or contribution. If there is only one the appendix is designated ``Appendix'', or ``Appendix 1'', or ``Appendix 2'', etc. if there is more than one.

%\begin{equation}
%a \times b = c
%\end{equation}

\end{document}